# Evolution of electronic and crystal structure during room-temperature annealing of quenched $REBa_2Cu_3O_{6+\delta}$, RE=Y, Nd

A.V. Fetisov, S.Kh. Estemirova

*Institute of Metallurgy, Ural Branch of the Russian Academy of Science, 101 Amundsen Str., Ekaterinburg, Russia*

**Abstract:** The crystal structure parameters and the low-temperature magnetic moment of the HTSC cuprates $YBa_2Cu_3O_{6+\delta}$, $Y_{1-x}Ca_xBa_2Cu_3O_{6+\delta}$, and $Nd_{1+x}Ba_{2-x}Cu_3O_{6+\delta}$, $x = 0.2$ are recorded after quenching them from the temperature of oxidative annealing. The subject of the present study is the aging effect, which results in an increase in the critical temperature $T_c$ and a decrease in the $c$ parameter of the crystal lattice for a while after the quenching. On the example of $YBa_2Cu_3O_{6+\delta}$, it is shown that the oxygen content dependence of $c$ undergoes the following changes with time: (1) there is an increase in the dependence slope with respect to the $(6+\delta)$-axis; (2) there is an increase in the dependence nonlinearity. The first type of changes is explained by an increase in the valence of copper ions in the $CuO_\delta$ planes that is accompanied by a decrease in their radius. The second type is explained by the electrostatic interaction of the $CuO_2$ structural planes due to the accumulation on them electron holes. Calculation of the parameter $c$ changes shows good quantitative agreement with the experimental data. The data obtained for the compounds $Y_{1-x}Ca_xBa_2Cu_3O_{6+\delta}$, and $Nd_{1+x}Ba_{2-x}Cu_3O_{6+\delta}$ (the cases of the hole and electron doping of $CuO_2$-planes) are discussed on the basis of the model representation developed by us for $YBa_2Cu_3O_{6+\delta}$.

Upon annealing to room temperature (RT), the quenched cuprate $RBa_2Cu_3O_{6+\delta}$ (R = Rare-earth elements or Y), which is a high-temperature superconductor, changes some of its properties, primarily the lattice parameter $c$ and the critical temperature $T_c$ [1–3]. One of examples is an increase in $T_c$ from 0 to 20 K and a decrease in the parameter $c$ by 0.04%, observed for the $YBa_2Cu_3O_{6.41}$ sample for several days after quenching [1]. Typically, this effect (*aging effect*) is considered to be associated with the oxygen ion ordering in the basic $CuO_\delta$ plane, in the result of which the defect-free copper-oxygen chains are formed [1–3]. This ordering leads to an increase in the concentration of hole charge carriers in both the $CuO_\delta$ and the neighboring $CuO_2$ planes and, as a consequence, to an increase in $T_c$. Meanwhile, the existence of a direct connection between the oxygen ordering and the concentration of holes in $RBa_2Cu_3O_{6+\delta}$ is not an experimentally confirmed fact. Moreover, a detailed analysis of publications on this subject in [4] allowed concluding that the arguments in favor of such a connection are groundless. Solving this problem is hindered by the absence at the present time of a reliable method for determining the concentration of the hole-charge in the $CuO_2$ planes ($q$) where high-temperature superconductivity is realized. Existing methods: the calculation of bond valence sums (BVS) for $CuO_2$-plane atoms [5] and the use of an empirical relationship between $q$ and thermoelectric power measured at RT [6] cannot be considered reliable for a number of reasons. For instance, the thermoelectric power of $RBa_2Cu_3O_{6+\delta}$ is a function of the state not only of the $CuO_2$ structural planes, but also of $CuO_\delta$ having metallic conductivity. In turn, the BVS method requires detailed structural analysis, which is difficult to provide when examining, for example, the initial stage of the aging process.

In this paper, we propose a simple original method for determining the type and concentration of free charge carriers in the $CuO_2$ structural planes of layered cuprates $RBa_2Cu_3O_{6+\delta}$ by analyzing the nonlinear form of the experimental dependences $c(\delta)$. With the help of this method it is shown that oxygen orderings occurring in $RBa_2Cu_3O_{6+\delta}$ at various $\delta$ do not lead to appreciable changes in the value of $q$, confirming the point made in [5]. It is also shown that heterovalent substitution on cation sites adjacent to the $CuO_2$ plane leads to an effect similar to the well-known "anomaly 1/8" [7–9] – a sharp drop in $T_c$ within a narrow region of $\delta$ corresponding to the hole concentration in the $CuO_2$ plane equal to 1/8.

**Experimental procedure**

Solid solutions $YBa_2Cu_3O_{6+\delta}$, $Y_{1-x}Ca_xBa_2Cu_3O_{6+\delta}$ and $Nd_{1+x}Ba_{2-x}Cu_3O_{6+\delta}$, $x = 0.2$ were synthesized from the oxides $Y_2O_3$, $Nd_2O_3$, CuO and barium carbonate at 950°C–100 h (R = Y) or 1000°C–48 h (R = Nd). As the result, a single phase product of the tetragonal structural modification was formed. To obtain samples with different oxygen content, portions of the synthesized material (by ~5–6 g) were oxidized at different temperatures from the interval 470–940 °C in air atmosphere and then quenched. The oxygen content was monitored by the change in mass of a small portion of a sample during additional 1 h oxidation at 470 °C in air. Using the values of $\Delta m$, the initial composition of a sample was determined by the formula:

$$\delta = \delta^{470} - \left[\frac{2.242 \cdot m_s}{\Delta m}\right], \tag{1}$$

where $\delta^{470}$ is the oxygen content in $RBa_2Cu_3O_{6+\delta}$ corresponding to the conditions of its additional oxidative ($t = 470$ °C, $p_{O_2} = 21$ kPa, $\tau = 1$ h): 6.90 (R = Y) и 6.95 (R = Nd) [10, 11]; $m_s$ is mass of the sample.

Every sample quenched after oxidative annealing was examined by X-ray diffraction (a Shimadzu XRD-7000 diffractometer) according to the scheme of the survey: (1) in 0.30±0.03 h after quenching (hereafter $\tau = 0.3$ h), (2) in 5.1 h ($\tau = 5$ h), (3) in 15 days ($\tau = 15$ days). The middle of the survey duration was taken for the calculation of $\tau$. Structural characteristics of the sample's material were determined by analyzing XRD peaks in the range $2\Theta = 20$–$70°$.

The low-temperature magnetic measurements were performed in a vibrating sample magnetometer (VSM) Cryogenic CFS-9T-CVTI, measuring the in-field cooling magnetization at 50 Oe in the temperature range 4.2–150 K with a rate of 1 °C/min.

**Experimental results**

Figure 1 shows the oxygen content dependences of lattice parameters $a$, $b$, and $c$ obtained through different time intervals $\tau$.

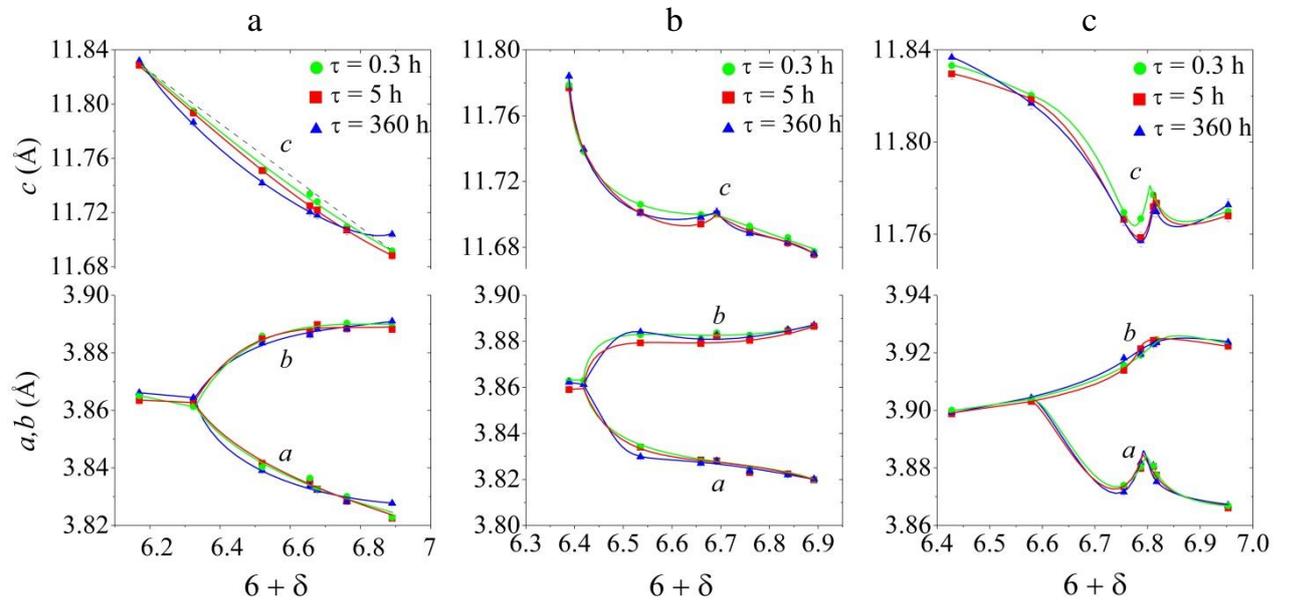

**Fig. 1.** Experimental dependences of lattice parameters $c$, $a$, and $b$ on oxygen content ($6 + \delta$) and the post-quench time ($\tau$) for: $YBa_2Cu_3O_{6+\delta}$ (a); $Y_{0.8}Ca_{0.2}Ba_2Cu_3O_{6+\delta}$ (b) и $Nd_{1.2}Ba_{1.8}Cu_3O_{6+\delta}$ (c). The curves are guides to the eye.

The course of the $a$ and $b$ ($\delta$) curves reflects the existence of the structural transition between tetragonal and orthorhombic phases, occurring at (6 + $\delta$): 6.33 ($YBa_2Cu_3O_{6+\delta}$); 6.42 ($Y_{0.8}Ca_{0.2}Ba_2Cu_3O_{6+\delta}$) and 6.59 ($Nd_{1.2}Ba_{1.8}Cu_3O_{6+\delta}$). With increasing $\tau$, the $c(\delta)$ dependence of $YBa_2Cu_3O_{6+\delta}$ transforms from linear at $\tau\to 0$ to nonlinear with growing nonlinear part. The $c(\delta)$ dependences of the samples with heterovalent substitution are already nonlinear at $\tau\to 0$ and for all $\tau$ have V-shaped singularities at (6 + $\delta$): 6.7 ($Y_{0.8}Ca_{0.2}Ba_2Cu_3O_{6+\delta}$); 6.8 ($Nd_{1.2}Ba_{1.8}Cu_3O_{6+\delta}$).

The results of the low-temperature magnetic study of the samples stored at RT for 360 h are presented in Fig. 2. It should be noted that the 360-h exposition of the samples after quenching ought to be considered as an approximation to the limit $\tau\to\infty$, since the parameters $c$ and $T_c$, as experienced by many researchers, are not farther modified.

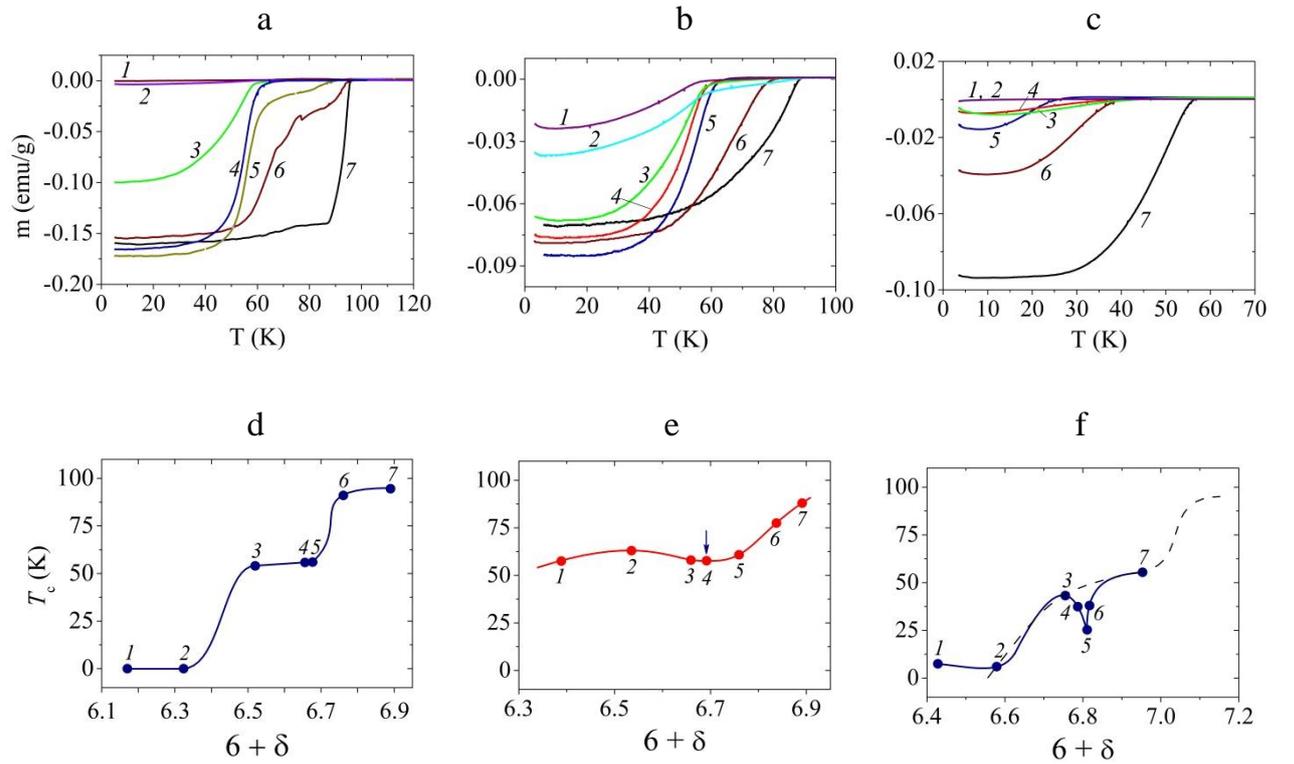

**Fig. 2.** Low-temperature magnetic studies: (a–c) temperature dependence of the magnetic moment of the samples; (d–f) superconducting transition temperature $T_c^{onset}$ (the beginning of the superconducting transition) as a function of oxygen content. The samples under investigation are: $YBa_2Cu_3O_{6+\delta}$ (a, d); $Y_{0.8}Ca_{0.2}Ba_2Cu_3O_{6+\delta}$ (b, e) and $Nd_{1.2}Ba_{1.8}Cu_3O_{6+\delta}$ (c, f). Arrow indicates a local diminishing of $T_c$ near $\delta = 0.7$ for the $Y_{0.8}Ca_{0.2}Ba_2Cu_3O_{6+\delta}$ sample.

It can be seen that the $T_c(\delta)$ dependences found for the samples under study are either a stepwise dependence with two "plateaus" at 55 and 90 K (for $YBa_2Cu_3O_{6+\delta}$), or they are fragments of stepwise dependencies, which is vividly manifested in the case of $Nd_{1.2}Ba_{1.8}Cu_3O_{6+\delta}$ cuprate (see the experimental curve and the extrapolating dashed line in Fig. 2f). The $T_c(\delta)$ dependences obtained by other authors for the unsubstituted compounds $YBa_2Cu_3O_{6+\delta}$ [12], и $NdBa_2Cu_3O_{6+\delta}$ [13] are similar to that in Fig. 2d and the dotted curve in Fig. 2f. Thus, heterovalent substitutions in these cuprate compounds result in a shift of the $T_c(\delta)$ dependencies toward higher $\delta$ in the case of substitution of Ca for Y and in the opposite direction in the case of substitution of Nd for Ba.

Fig. 2 also shows that in the vicinity of compositions $(6 + \delta)$: 6.7 for $Y_{1-x}Ca_xBa_2Cu_3O_{6+\delta}$ and 6.8 for $Nd_{1+x}Ba_{2-x}Cu_3O_{6+\delta}$ a local decrease in the critical temperature is observed. These ranges of $\delta$ coincide with the corresponding regions of the increase in the lattice parameter $c$ (besides, in the parameter $a$ for the neodymium composition).

**Discussion**

The local decrease in the critical temperature occurring in substituted cuprates at certain values of $\delta$ (see Fig. 2e and 2f) is very similar to the suppression of $T_c$ in some others cuprates [7–9] when the concentration of charge carriers reaches 1/8 (for example, in $La_{2-x}Sr_xCuO_4$, a local drop of $T_c$ by ~20 K occurs at $x = 0.21$ [7]). At the given charge carrier concentration so-called *charge stripes* arise. They are a static ordering of charge in the form of alternating regions with an increased and lowered charge density. Then, the integral $T_c$ of the superconductor is specified by the regions with the worst superconducting characteristic (having reduced charge density). Although in $RBa_2Cu_3O_{6+\delta}$ a notable local drop in $T_c$ has not been observed so far, it is considered [14] that these cuprates are thermodynamically "close" to the state with static charge stripes. At least, in $RBa_2Cu_3O_{6+\delta}$ dynamic stripe correlations were found, which could become static in the presence of sufficiently effective pinning centers [9, 15]. In our opinion, the heterovalent substitution in $RBa_2Cu_3O_{6+\delta}$ in the sites adjacent to the $CuO_2$ plane results in charge inhomogeneity that could be translated to the $CuO_2$ plane, originating charge stripes at certain value of $\delta$.

Now let us consider in more detail the $\delta$-dependences of the parameter $c$ obtained for the pure cuprate $YBa_2Cu_3O_{6+\delta}$ (see Fig. 1a). In this case, an obvious regularity is observed: the degree of nonlinearity of the curves obtained increases with the time elapsed after quenching of the samples. In Fig. 3 symbols represent the extracted nonlinear part of the experimental dependences $c(\delta)$ from Fig. 1a. Our main assumption is that the dependences presented in Fig. 3 directly reflect the process of the hole charge transfer from the $CuO_\delta$ structural plane to $CuO_2$, which occurs while the oxygenation and/or aging processes are in progress in $YBa_2Cu_3O_{6+\delta}$. To prove this, we calculate the nonlinear part $\Delta c^*(\delta)$ of the curve $c(\delta)|_{\tau=360h}$, using the electrostatic equation for interacting charged planes, Fig. 4. Forces (per unit area) arising as a result of charge transfer between $CuO_\delta$ and $CuO_2$ structural planes can in general be represented as follows:

$$F_1 = \sum_n B_n \frac{(q \cdot p_n)\bar{e}^2}{2\varepsilon_0 \varepsilon^n \cdot \sigma^2} + \frac{\bar{e}^2}{2\varepsilon_0 \sigma^2}\left[\frac{(1-2q)q}{\varepsilon^{III}} + \frac{q^2}{\varepsilon^{II}}\right]; \qquad (2a)$$

$$F_2 = \sum_m B_m \frac{(q \cdot p_m)\bar{e}^2}{2\varepsilon_0 \varepsilon^m \cdot \sigma^2} + \frac{q^2 \bar{e}^2}{2\varepsilon_0 \varepsilon^I \sigma^2}, \qquad (2б)$$

where $p_{n,m}$ is the dimensionless charge (per one copper ion) that may exist in structural planes different of $CuO_\delta$ and $CuO_2$ (here the summation of effects of these planes to the $CuO_2$ planes is executed); $\varepsilon_0$ is the vacuum permittivity; $\varepsilon^n$, $\varepsilon^m$ are the dielectric permittivity of the matter placed between two structural planes under consideration.

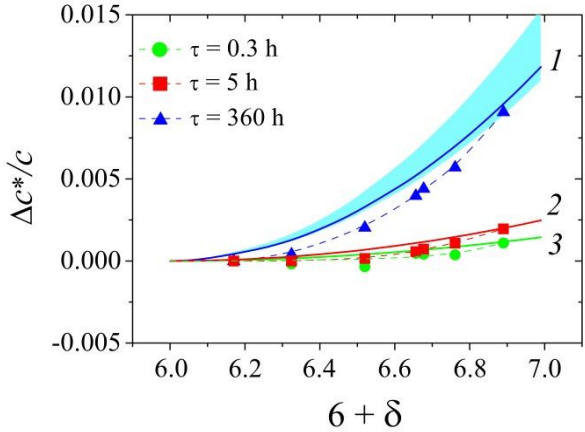
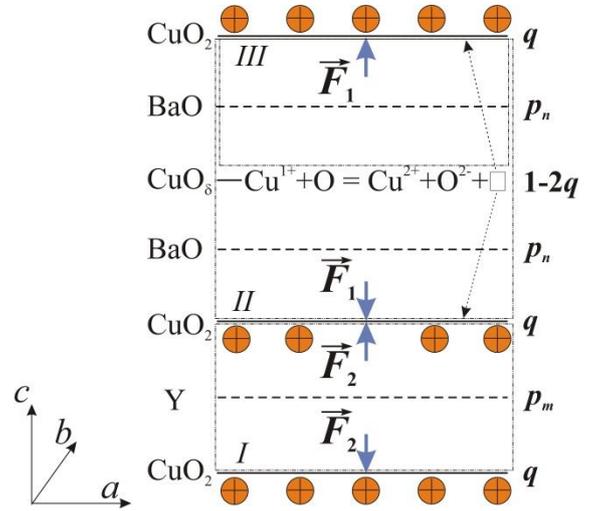

**Fig. 3.** Extracted nonlinear part of the $\Delta c(\delta)$ experimental dependences for $YBa_2Cu_3O_{6+\delta}$. Dots and connecting lines concern the experimental data, solid lines are the calculation according to equations (2) and (3). The interval of uncertainty for the calculated curve $\tau = 360$ h is shown by shading.

**Fig. 4.** There is shown the electrostatic forces $F_1$, $F_2$ acting on the $CuO_2$-structural planes in the frames of the elementary cell of $YBa_2Cu_3O_{6+\delta}$. On the right side of the structural planes the appropriate charge is given. The reaction resulting in holes in the $CuO_\delta$ planes is represented.

In Fig. 4, three spatial regions between one $CuO_\delta$ and two $CuO_2$ planes are depicted in the contours designated I, II and III. These sections correspond to permeability $\varepsilon^I$, $\varepsilon^{II}$ and $\varepsilon^{III}$; σ is an elementary area (σ = a · b), to which the specific charges $q$, $p_n$ and $p_m$ relate. The relationship between $q$ and δ can be represented as $q \approx 0.187\cdot\delta$ [6], which has been obtained experimentally with the help of the BVS method. The quadratic term in equations (2a) and (2b) describes the direct interaction between the $CuO_\delta$ and $CuO_2$ structural planes, which is a result of the charge transfer process, and the linear terms of the equations – the interactions of these planes with all other structural elements of the unit cell with non-null but independent of δ electric charge. For the purposes pursued in the present paper we shall consider only the nonlinear terms in equations (2a) and (2b):

$$F'_1 = \frac{q^2\bar{e}^2}{2\varepsilon_0\varepsilon^{II}\sigma^2} - \frac{2q^2\bar{e}^2}{2\varepsilon_0\sigma^2\varepsilon^{III}}; \qquad (3a)$$

$$F'_2 = \frac{q^2\bar{e}^2}{2\varepsilon_0\varepsilon^I\sigma^2}. \qquad (3б)$$

Then, the relative change of the lattice parameter $c$ induced by the electrostatic interaction of structural plates can be written as:

$$\frac{\Delta c}{c} = \frac{\frac{2}{3}F'_1 + \frac{1}{3}F'_2}{\varepsilon_{33}}, \qquad (4)$$

where $\varepsilon_{33}$ is the material elastic constant for the $c$ direction. With the values $\varepsilon_0 = 8.854\cdot10^{-12}$ C²/(N·m²); $\varepsilon^I = \varepsilon^{II} = 1$ [\*]; $\varepsilon^{III} = 4$ [\*\*]; $\sigma^2 = 1.5\cdot10^{-38}$ m⁴; $\bar{e} = 1.6\cdot10^{-19}$ C; $\varepsilon_{33} = 186$ GPa [17] [\*\*\*] and $q = 0.187\cdot\delta$ in equations (3) and (4) we obtain the theoretical de-

---

[\*] This value accepted for $\varepsilon^I$ and $\varepsilon^{II}$ we explain by the fact that inside the closed contours I, II, which are marked in Fig. 4 by dot-dashed line, the electric field strength of the charges in $CuO_2$ planes, according to the of electrostatic laws, equals zero. Consequently, there must not be any reaction of the substance to the appearance of charges in $CuO_2$.

[\*\*] On the other hand, between the planes $CuO_2$ and $CuO_\delta$ (inside the contour III) there are "ionic" layers $Ba^{2+}O^{2-}$, which are easily polarized under the action of the electric field from these planes [16]. Then, the parameter $\varepsilon^{III}$ can be estimated by the dielectric permittivity that is characteristic of ionic compounds: $\varepsilon^{III} \approx 4$–10. It should be noted that the specified range of the parameter $\varepsilon^{III}$ affects the resulting value $\Delta c/c$ negligible. The calculation has shown that as $\varepsilon^{III}$ increases between the specified range limits, $\Delta c/c$ goes up by only 30%.

[\*\*\*] According to [18, 19], there is reason to believe that the structural unit enclosed in contour II (see Fig. 4) is harder than the block in contour I (hereinafter – structural blocks II and I). We cannot take this difference into account without having specific data. However, if we consider the limit case accepting the elastic constant of the block I as ∞ (then, the elastic constant of the block II must be equal to 2/3 of the integral value $\varepsilon_{33}$ determined experimentally), then it is not difficult to show that the wanted value $\Delta c/c$ decreases only by 7%.

pendence $\Delta c/c$ that turns out to match the experimental data for the case $\tau = 360$ h quite well (see curve *1* in Fig. 3). The error interval related to both the uncertainty of the value of $\varepsilon^{III}$ and the possible difference in the elastic constant of the structural blocks I and II (see Fig. 4) is shown in Fig. 3 by toning. A good quantitative agreement between the calculated curve and the experimental data gives us reason to believe that our basic assumption is correct. Then, the experimental dependences in Fig. 3 obtained at $\tau = 0.3$ and 5 h reflect the situation when the charge transfer from the $CuO_\delta$ plane to $CuO_2$ is not fully realized. These dependencies are described by the above equations, in which charge $q$ is multiplied by factors of 0.32 and 0.48 for $\tau = 0.3$ and 5 h, respectively (curves *3* and *2* in Fig. 3).

Thus, we found that the nonlinearity of $c(\delta)$ dependences for $YBa_2Cu_3O_{6+\delta}$ is directly related to the concentration of charge carriers (in this case, holes) in the $CuO_2$ structural planes, and the extracted nonlinear part of these dependences is proportional to the square of the charge concentration. Based on this thesis, we further can analyze the $c(\delta)$ dependences obtained for substituted $RBa_2Cu_3O_{6+\delta}$ (see Figs. 1b and 1c). For some reasons, which we are not ready to discuss now, they are essentially nonlinear right after quenching. This initial nonlinearity may not be related to the charge accumulated on the $CuO_2$ planes. At the same time, further we show that such dependencies can also yield useful information. In this case, we will analyze not the dependences themselves, but the changes that occur to them with time after quenching. Thus, the nonlinear part will be extracted from the differential dependences $\Delta c/c$ ($\delta$), assuming that it is nonlinearity that is related to the electrostatic interaction of the $CuO_2$ planes.

Fig. 5 represents differential dependences $\Delta c/c$ ($\delta$) obtained from the initial experimental curves $c(\delta)$ by subtracting the $c(\delta)$ curve for $\tau = 0.3$ h and dividing by $c$. The extracted nonlinear parts ($\Delta c^*/c$) of that dependences are shown in Fig. 6. Let us consider within the framework of the present discussion these nonlinear parts without quantitative analysis, purely qualitative. For $Y_{1-x}Ca_xBa_2Cu_3O_{6+\delta}$ the degree of nonlinearity of the $\Delta c^*/c$ ($\delta$) curve gradually increases in time (see Fig. 6a), which does not differ from the behavior of the unsubstituted yttrium-barium cuprate. It is easily explained by the hole doping of the $CuO_2$ planes occurring without any charge redistribution between planes when Y is substituted with Ca (i.e., the $CuO_2$ planes are not get charged during such dop-ing) and there are no other differences from the electronic structure of $YBa_2Cu_3O_{6+\delta}$ [6, 20]. At the same time the $\Delta c^*/c$ ($\delta$) dependence obtained for $Nd_{1+x}Ba_{2-x}Cu_3O_{6+\delta}$ varies quite differently with time $\tau$ (see Fig. 6b). At the initial stage the $\Delta c^*/c$ ($\delta$) dependence is bent downward and only then one acquires the usual dynamics with bending upward. This behavior logically follows from the assumption that when $Ba^{2+}$ is replaced by $Nd^{3+}$, the $CuO_2$ planes are doped by electrons with the charge redistribution between BaO and $CuO_2$ (see Fig. 4). Then, a repulsive interaction of the $CuO_2$-structural planes would occur at the initial stage of aging. At this case it would be due to the negative charge localized in $CuO_2$. However, as the electron holes enter these planes, the total charge in these planes at the beginning decreases over time and only then, after the recombination of all electrons with electron holes, begins to increase.

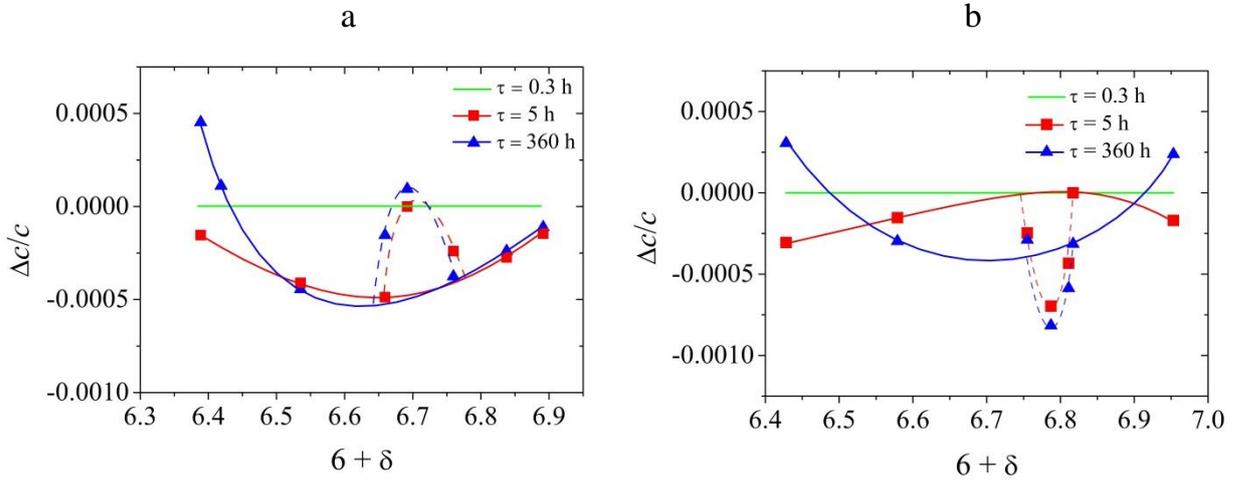

**Fig. 5.** Relative change of the lattice parameter $c$ corresponding to aging of $Y_{0.8}Ca_{0.2}Ba_2Cu_3O_{6+\delta}$ (a) and $Nd_{1.2}Ba_{1.8}Cu_3O_{6+\delta}$ (b) from $\tau = 0.3$ h to 5 and 360 h. The values are obtained by the subtraction of the experimental $c|_{\tau=0.3h}$ from $c|_{\tau=5h}$ or $c|_{360 h}$.

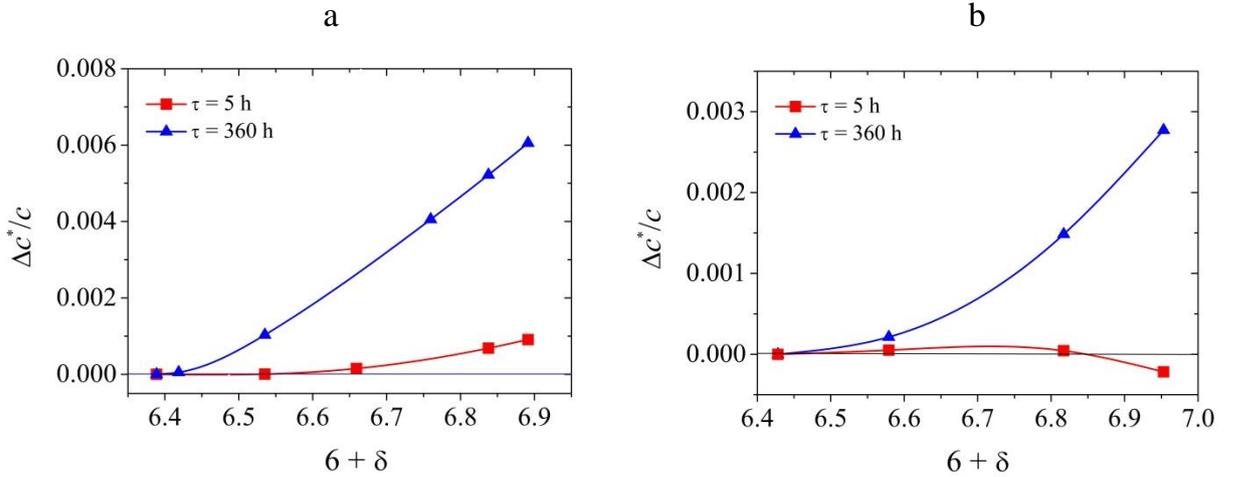

**Fig. 6.** Extracted nonlinear part of the $\Delta c/c$ ($\delta$) dependences for $Y_{0.8}Ca_{0.2}Ba_2Cu_3O_{6+\delta}$ (a) and $Nd_{1.2}Ba_{1.8}Cu_3O_{6+\delta}$ (b), corresponding to aging time 5 and 360 h.

On the other hand, however, there is a rule of thumb related to the change of charge carrier type in the electronic system of layered cuprates: such a change cannot occur without substantial rearrangement of crystal lattice [21]. This rule is based on unsuccessful attempts to synthesize a layered cuprate containing $CuO_2$ layers which have *n*-type conductivity. Based on these data, it is generally accepted that when Ba is replaced by neodymium, an extra electron does not dope the $CuO_2$ planes in $Nd_{1+x}Ba_{2-x}Cu_3O_{6+\delta}$ but forms a bond with additional oxygen localized in the $CuO_\delta$ planes. Indeed, the equilibrium amount of oxygen in $Nd_{1+x}Ba_{2-x}Cu_3O_{6+\delta}$, according to different data, exceeds that for $NdBa_2Cu_3O_{6+\delta}$. At the same time, an exception for the above rule was recently found: in the course of the saturation of $Y_{0.38}La_{0.62}(Ba_{0.87}La_{0.13})Cu_3O_y$ with oxygen the change in the type of conductivity from *n* to *p* was recorded without a noticeable change in the crystal structure [22]. This solid solution is a close analogue of $Nd_{1+x}Ba_{2-x}Cu_3O_{6+\delta}$. According to the above qualitative analysis, heterovalent substitution in $Nd_{1+x}Ba_{2-x}Cu_3O_{6+\delta}$ leads to doping of this cuprate by electrons at least partially, and then, with the cuprate being saturated with oxygen, a "smooth" change in the type of charge carriers occurs. Thus, $Nd_{1+x}Ba_{2-x}Cu_3O_{6+\delta}$ appears to be another exception to the empirical rule proposed by the authors [21].

In conclusion we draw attention that there are no any peculiarities (breaks) in the experimental curves $\Delta c/c^*$ ($\delta$) in Figs. 3 and 6 at $\delta$ corresponding to the characteristic transitions in $RBa_2Cu_3O_{6+\delta}$ (ordering/disordering of oxygen in a $CuO_\delta$ plane): tetra → ortho-II, ortho-II → ortho-III, and ortho-III → ortho-I, which for $YBa_2Cu_3O_{6+\delta}$ are observed at $\delta = 0.32$, 0.63, and 0.77, respectively [23–25]. Consequently, oxygen orderings occurring in $RBa_2Cu_3O_{6+\delta}$ do not affect the charge concentration in $CuO_2$, and the assumption of the authors of Refs. 1–3, that the aging effect in $RBa_2Cu_3O_{6+\delta}$ is related to the oxygen ordering in $CuO_\delta$, is apparently not true. In our opinion, the aging of these cuprates is associated with kinetic difficulties for the process of charge transfer from the $CuO_\delta$ planes to $CuO_2$, occurring over a rather long chain: $O_{CuO_\delta} \to Cu_{CuO_\delta} \to O_{ap} \to Cu_{CuO_2} \to O_{CuO_2}$ (where $O_{ap}$ is apical oxygen connecting the planes $CuO_\delta$ and $CuO_2$).

**Acknowledgements**

The authors are grateful to Dr. V.Ya. Mitrofanov, Dr. S.A. Uporov, and Dr. G.A. Dorogina† for carrying out the magnetic measurements. The present work was performed in accordance with the State Program of IMET UrD RAN [grant no. 0396-2015-0075], with using the equipment of the Resource Center "Ural-M".